\documentclass[nonacm]{acmart}
\def \bs {\boldsymbol{s}}
\def \balpha {\boldsymbol{\alpha}}
\usepackage[super]{nth}
\AtBeginDocument{%
  \providecommand\BibTeX{{%
    \normalfont B\kern-0.5em{\scshape i\kern-0.25em b}\kern-0.8em\TeX}}}

\begin{document}
%\settopmatter{authorsperrow=4}
\title{Multi-objective Ranking via Constrained Optimization}

\author{Michinari Momma}
\affiliation{Amazon}
\email{michi@amazon.com}

\author{Alireza Bagheri Garakani}
\affiliation{Amazon}
\email{alirezg@amazon.com}

\author{Nanxun Ma}
\affiliation{University of Washington}
\email{nanxunma@uw.edu}

\author{Yi Sun}
\affiliation{Amazon}
\email{yisun@amazon.com}

\renewcommand{\shortauthors}{Michinari Momma, et al.}

\begin{abstract}
In this paper, we introduce an Augmented Lagrangian based method to incorporate the multiple
objectives (MO) in a search ranking algorithm. Optimizing
MOs is an essential and realistic requirement for building ranking models in production.
The proposed method formulates MO in constrained optimization and
solves the problem in the popular Boosting framework -- a novel
contribution of our work. Furthermore, we propose a procedure to set
up all optimization parameters in the problem.
The experimental results show that the method
successfully achieves MO criteria much more
efficiently than existing methods.
\end{abstract}

%  \begin{CCSXML}
% <concept_id>10002951.10003317.10003338.10003343</concept_id>
% <concept_desc>Information systems~Learning to rank</concept_desc>


%\ccsdesc[500]{Computer systems organization~Embedded systems}


\maketitle
\section{Introduction} \label{sec:intro}
In the real production environment, search relevance modeling faces
unique challenges; Due to the multi-dimensional nature of relevance, use of single
objective does not suffice to capture the concept. For example, in
product search, customer response such as purchase, etc., are
used as a target to optimize
\cite{Sorokina:2016:ASJ:2911451.2926725}. However, such a target may
not represent important concepts such as customer engagement,
membership benefit, product quality and defects\footnote{search results that do not match the query in
various aspects}, etc. 
Moreover, business constraints are additional requirements in production
modeling; Some are derived from existing relevance metrics proven to be
effective over time. Others are from operational and strategic requirements. Examples
include latency, minimum \%-gain to consider launch
and avoiding adult items to surface, etc. All of these requirements need to be satisfied in production modeling.

Traditional machine learning ranking solutions such as
$\lambda$-MART\cite{from-ranknet-to-lambdarank-to-lambdamart-an-overview}
cannot handle such complicated objectives in a systematic manner.
Instead, multiple objective (MO) optimization \cite{Kais99} should be leveraged to
provide a robust and scalable way to model MO problems.
To design a solution
in MO, it is important to have clear requirements. Production modeling
often has specific goals such as achieving \%-gain over baseline models, or satisfying specific
business constraints. Such goals are clearly determined among stakeholders. To handle such requirements, we propose a
constrained optimization approach and leverage constraints to formulate the requirements.
Further, we develop a MO methodology as an extension to Boosting /
$\lambda$-MART, which is a popular and standard approach in ranking
modeling. Many practitioners / companies using it can immediately
leverage our method for their production modeling.

Challenges in formulating constrained optimization in $\lambda$-MART for production modeling includes 1) optimization done
over the function space where the function evaluation is costly and 2) the number
of iterations (i.e., \#trees) is limited due to the latency requirement.
To alleviate them, adaptation of the
Augmented Lagrangian (AL) method \cite{nocedal2006numerical} to
$\lambda$-MART (AL-LM) is proposed. 
AL allows us to solve the constrained optimization by iteratively
solving unconstrained problem (i.e., AL).
With AL, we can solve the constrained optimization problem
by jointly optimizing both dual and primal (i.e., Boosting).
To the best of our knowledge, our work is the first to explicitly introduce constrained optimization problem in Boosting and the first to apply it search relevance problems.
To use AL-LM in modeling, we propose the ``one shot modeling'' 
where the MO model is built with only a few trials after constraint
parameters are found.
The performance of AL-LM has been validated on both public ranking data et as well as online production systems.
\section{Augmented Lagrangian in Boosting}
Suppose our goal is to optimize (T-) multiple metrics on ranking and each
metric is measured in the normalized discounted cumulative gain
(NDCG). The minimum criteria to achieve for each objective is given as
upper bounds (UB) in the cost function. Specifically, we employ the same surrogate cost function on NDCG used
in $\lambda$-MART and set UB on the cost $b^t$ (i.e., $C^t
 (\boldsymbol{s}) \le b^t, \ t=1,…,T$, with $\bs$ being the predictive scores of the model).
 Usually, we set UB as fraction (\%) of the cost of a certain baseline
 model. Therefore, we rescale the cost accordingly, so that UB is very intuitive; setting $b=0.9$ implies cost reduction by 10\%. 
Given the constraints represented in terms of cost functions, we have the following constraint
optimization problem:
$\min_{\boldsymbol{s}} C^{pm}\left(\boldsymbol s\right) \ s.t. \ {C}^t
\left( \boldsymbol{s}\right) \le {b}^t, \ t =1,...,T  , ~ pm:
{\rm primary ~ objective}
%\end{equation}
$.
With the dual variables $\balpha$, AL at iteration $k$ is written
as follows:
$$
\mathcal{L}_k \left( \bs, \balpha \right) 
= C^{pm} \left( \bs \right)  
+ \sum_t^T 
\alpha^t \left( {C}^t \left( \bs  \right)  - {b}^t \right) 
- \sum_t^T \frac {1} {2 \mu} 
\left( \alpha^t -
    \alpha_{k-1}^t \right)^2 
 $$ 
where $\alpha_{k-1}^t$ is a solution in the previous iteration and a constant in
the current iteration $k$. $\mu$ is a sufficiently large constant\footnote{$\mu=10$ is
large enough for datasets we used, hence the value 10 is used for all cases.}.
Note that the last term is the augmented term and it gives proximal
minimization with iterates $\alpha_{k-1}^t$, to make the Lagrangian optimization
{\it smooth}.

We maximize the Lagrangian with respect to $\balpha \ge 0$ and
minimize with respect to $\bs$:
$
\max_{\balpha \ge 0} \min_{\bs} \mathcal{L}_k \left( \bs , \balpha \right)
$.
From the stationary condition ${\partial \mathcal{L}_k} / {\partial  \alpha^t} =0$, 
we obtain the update formula for $\alpha$:
$$
\alpha_k^t = \max \left(0,   \mu ( {C}^t \left(\bs \right) - {b}^t )
  + \alpha_{k-1}^t \right)
$$
At an iteration $k$, if the constraint $t$ is not satisfied, i.e., $C^t
\left( \bs\right) > b^t$, we have $\alpha_k^t > \alpha_{k-1}^t$, which means 
$\alpha^t$ increases unless the constraint is already satisfied --
focusing more on unsatisfied constraints during the optimization iterations.
As for the primal, we leverage the gradient boosting tree framework where we plug in derivatives of AL into the algorithm in
\cite{from-ranknet-to-lambdarank-to-lambdamart-an-overview}. 
The algorithm of AL-LM looks very similar to that of $\lambda$-MART except 
update of $\balpha$ at each Boosting iteration. Thus, the modification to existing
solvers should require a minimal effort.
\\
\\
\textbf{One shot modeling to leverage AL-LM}:
One requirement to set up AL-LM is to find a right UB associated with
the goal given by a metrics (i.e., NDCG). To find such UB values, we propose
the following 3-step prodecure; 
1) run a unconstrained model to build a baseline and 
obtain the cost value for each sub-objective.
2) identify the UB for each sub-objective independently by running 1D search on UB
values.
This step conducts the sensitivity analysis -- building models with the primary and a
single sub-objective, and find UB that has good balance between them.
Note this step should be run in parallel for all sub-objectives to
gain efficiency.
Note we need to look at validation results to avoid overfitting.
3) apply all the UB values identified in the step 2 and
build a model with the full set of constraints.
\section{Experiments} \label{sec:experiment}
Here, we show steps of MO modeling using MSLR-10K
dataset\cite{DBLP:journals/corr/QinL13}.  Then, we apply the
methodology to our proprietary product search dataset to illustrate
how a production modeling is done.
\\\\
\textbf{MO model building using MSLR dataset}:
To build MO models with the MSLR dataset, we use the relevance
judgement as the primary objective and the following 5 features as
sub-objectives\footnote{as QS/QS2 are
  badness score, we linearly convert the features to goodness score}:
\textit{QualityScore} (QS), \textit{QualityScore2} (QS2),
\textit{PageRank} (PR),
\textit{UrlClick} (UC) and
\textit{UrlDwellTime} (UDT).
To provide a case study of MO modeling, we define the modeling goal as follows:
{\em improve sub-objective NDCG, measured by
\%-gain from baseline, as much as possible while keeping the impact to
the relevance target by
-1\%, measured also as \%-gain from baseline.}

Due to space limitation, we cannot show results of step 1 and 2. After
running the unconstrained model in step 1, step 2 is done to choose UB
to keep -1\% goal in the primary objective by some margin, as the full
model will generally
degrade the value\footnote{step 2 choose (70, 50, 60, 80, 80)-\% as UB
  for PR, QS, QS2, UC and UDT, respectively}. 
Tab.\ref{tab:AL-LM-full} shows the result of full model (step
3).
The model achieves +1\% goals for all objectives
in the test set while keeping the relevance gain within -1\%. Notably, 
gains for PR, UC and UDT attain 10+\%.

Many existing methods such as \cite{learning-to-rank-with-multiple-objective-functions} use a linear weighting (LW) of
sub-objectives to build MO models. Here, we conduct a study
to compare performance of AL-LM with LW.
LW can be formulated as
$\min_{\boldsymbol{s}} w^{pm}C^{pm}(\bs) + \sum_t^T w^t{C}^t (\bs)$
with user-given weights: $w^{pm} + \sum_t w^t = 1$, which need to be
tuned to optimize MO.
To be comparable, the same cost functions / objectives are used to optimize.
We first run a similar exploration we did
for AL-LM (step 2) to gain efficiency.
After selecting promising subspaces, we build bunch of full
models by exploring combination of binary and random search in weights. 
After building 200+ full models, only one model is found to
satisfy all constraints. The best, and the only, result is shown in
\nth{2} row in Tab. \ref{tab:AL-LM-full}.  While the overall result is comparable, the number of model build is totally
different: AL-LM achieves the model with one trial after the UB setup
while LW model spends 200+ trials after the initial grid search.
\begin{table}[]
  \vspace{-3mm}
  \caption{Full model result for AL-LM and LW (\%-gain).}
  \vspace{-3mm}
\center
\scalebox{0.85}{
\begin{tabular}{|c|r|r|r|r|r|r|}
\hline
Model  & rel. & PR & QS & QS2 & UC & UDT \\ \hline
  AL-LM
& -0.82 & 10.38 & 1.16 & 1.26 & 10.87 & 12.06 \\ \hline
  LW
      & -0.89   & 4.83  & 1.00          & 1.34        &
                                                                       13.60 & 15.30 \\
\hline
\end{tabular}
}
  \vspace{-3mm}
\label{tab:AL-LM-full}
\end{table}
\\\\
\textbf{Product search modeling}:
Further, we apply AL-LM to our proprietary product search dataset.
The product search dataset consists of search queries, numerous input features as well
as customer’s purchase decision. We follow the basic modeling practice
described in \cite{Sorokina:2016:ASJ:2911451.2926725}.
The primary objective of this model building is to optimize NDCG of
purchased items. We have at least 4 sub-objectives such as reduction of search defect,
surfacing high quality products, etc.
We follow the
one-shot modeling procedure and tune the model to significantly better
in all sub-objectives while keeping the impact in the primary objective
insignificant. The offline results shows 2-4\% gain in sub-objectives
in 3 and significant gain in the rest, while keeping the impact to the
purchase objective insignificant.
The online A/B test confirmed the consistent behavior for all objectives. 
\section{Conclusion}
In this paper, we introduced AL-LM, a novel algorithm to implement
constrained optimization in Boosting. This allows us to build MO model
built on top of $\lambda$-MART. The experimental results showed AL-LM
successfully built MO models much more efficiently than existing
linear weighting methods.
\bibliographystyle{ACM-Reference-Format}
\bibliography{www-mo-final.bib}

\end{document}